\documentclass[11pt]{article}
\usepackage{epsfig}
\usepackage{threeparttable}
\usepackage{amsmath}
\usepackage{amsfonts}
\usepackage{amssymb}
\begin{document}
\newcommand{\apj}{{ApJ}}
\newcommand{\pasp}{{PASP}}
\newcommand{\aj}{{AJ}}
\newcommand{\apjl}{{ApJL}}
\newcommand{\aaps}{{A\&A S}}
\newcommand{\mnras}{{MNRAS}}
\newcommand{\apjs}{{ApJS}}
\newcommand{\aap}{{A\&A}}
\newcommand{\prd}{{Phys. Rev. D}}
\newcommand{\nat}{{Nature}}
\newcommand{\physrep}{{Phys. Rep.}}

\def\Ohat{{\widehat \Omega}}
\def\lsim{\,\lower2truept\hbox{${< \atop\hbox{\raise4truept\hbox{$\sim$}}}$}\,}
\def\gsim{\,\lower2truept\hbox{${> \atop\hbox{\raise4truept\hbox{$\sim$}}}$}\,}

\def\Omm{\Omega_m}
\def\Oml{\Omega_{\lambda}}
\def\Om0{\Omega_0}
\def\Omb{\Omega_b}
\def\Ombh{\Omega_b\thinspace h^2}

\def\di{\mbox{d}}
\def\Msun{M_{\odot}}
\def\Zsun{Z_{\odot}}

\def\simlt{\mathrel{\rlap{\lower 3pt\hbox{$\sim$}}\raise 2.0pt\hbox{$<$}}}
\def\simgt{\mathrel{\rlap{\lower 3pt\hbox{$\sim$}} \raise 2.0pt\hbox{$>$}}}

\newcommand{\nc}{\newcommand}
\newcommand{\beq}{\begin{equation}}
\newcommand{\eeq}{\end{equation}}
\newcommand{\be}{\begin{eqnarray}}
\newcommand{\ee}{\end{eqnarray}}
\newcommand{\Odm}{\Omega_{\rm dm}}
\newcommand{\Ob}{\Omega_{\rm b}}
\newcommand{\nb}{n_{\rm b}}
\newcommand{\num}{\nu_\mu}
\newcommand{\nue}{\nu_e}
\newcommand{\nut}{\nu_\tau}
\newcommand{\nus}{\nu_s}
\newcommand{\mnus}{M_s}
\newcommand{\taus}{\tau_{\nu_s}}
\newcommand{\nnt}{n_{\nu_\tau}}
\newcommand{\rnt}{\rho_{\nu_\tau}}
\newcommand{\mnt}{m_{\nu_\tau}}
\newcommand{\tnt}{\tau_{\nu_\tau}}
\newcommand{\bi}{\bibitem}
\newcommand{\rar}{\rightarrow}
\newcommand{\lar}{\leftarrow}
\newcommand{\lrar}{\leftrightarrow}
\newcommand{\dm}{\delta m^2}
\newcommand{\mpl}{m_{Pl}}
\newcommand{\mbh}{M_{BH}}
\newcommand{\nbh}{n_{BH}}
\def\zrec{z_{\rm rec}}
\def\zreio{z_{\rm reion}}
\def\kms{\ifmmode{{\rm km}\,{\rm s}^{-1}}\else{km\,s$^-1$}\fi}
\def\mpc{{\rm Mpc}}

\newcommand{\eq}{{\rm eq}}
\newcommand{\tot}{{\rm tot}}
\newcommand{\M}{{\rm M}}
\newcommand{\coll}{{\rm coll}}
\newcommand{\ann}{{\rm ann}}

\centerline{{\em \bf JENAM 2004 -- The many scales in the Universe}}

\vspace*{5.00 cm}

\centerline{\bf\Large{Cosmological reionization after WMAP:}} 
\vspace*{0.5 cm}
\centerline{\bf\Large{perspectives from {\sc PLANCK} and future CMB missions}}

\vspace*{2.00 cm}

\centerline{\em C.~Burigana$^1$, L.A.~Popa$^{1,2}$, F.~Finelli$^1$, 
R.~Salvaterra$^3$, G.~De~Zotti$^{4,3}$, and N.~Mandolesi$^1$} 

\vspace*{1.00 cm}

\centerline{$^1$INAF-CNR/IASF, Sezione di Bologna, via Gobetti 101, 
I-40129 Bologna (Italy)}
\centerline{$^2$Institute for Space Sciences, Bucharest-Magurele R-76900 (Romania)} 
\centerline{$^3$SISSA, via Beirut 4, I-34014 Trieste (Italy)}
\centerline{$^4$INAF - Osservatorio Astronomico di Padova, vicolo 
dell'Osservatorio 5, I-35122 Padova (Italy)}

\vspace*{1.00 cm}



\centerline{\bf Abstract} 

\vspace*{0.5cm}

\noindent 
The WMAP first year detection of a high redshift reionization through its 
imprints
on CMB anisotropy T and TE mode angular power spectra calls for a better  
comprehension of the universe ionization and thermal history after the 
standard recombination. Different reionization mechanisms predict 
different signatures in the CMB, both in temperature and polarization
anisotropies and in spectral distortions. The {\sc Planck} capability 
to distinguish among different scenarios through its sensitivity to
T, TE, and E mode angular power spectra is discussed. Perspectives 
open by future high sensitivity experiments on the CMB polarization
anisotropy and spectrum are also presented.

\section{Introduction}
\label{sec:introduction}

The accurate understanding of the ionization history of the universe
plays a fundamental role in modern cosmology.
The classical theory of hydrogen
recombination for pure baryonic cosmological models
\cite{peebles, zeldovich}, 
subsequently extended to non-baryonic dark matter models 
\cite{zabotin, jones, recfast, psh}
can be modified in various ways to take into account
additional sources of photon and energy production
able to significantly
increase the ionization fraction, $x_e$,
above the residual fraction ($\sim10^{-3}$)
after the standard recombination epoch
at $\zrec\simeq 10^3$.
These photon and energy production processes may leave imprints
on the cosmic microwave background (CMB) providing a crucial 
``integrated'' information on the so-called
{\it dark} and {\it dawn} {\it ages}, i.e. the epochs
before or at the beginning the formation of first stars and primeval galaxies,
complementary to those obtained by the study of the diffuse backgrounds
in other frequency bands and that are impossible or difficult to 
study with other direct astronomical observations.

Among the extraordinary results recently achieved by WMAP
\cite{bennett2003}
the detection of a cosmological reionization at relevant 
redshifts 
\cite{kogutetal03}
represents one of the new most remarkable
discovery in the recent years. 
In spite of the uncertainty associated to the degeneracy
\cite{bondetal_deg} between the Thomson scattering 
optical depth, $\tau$, and 
the spectral index of primordial perturbation, $n_s$,
in CMB anisotropy data,  
a cosmological reionization at substantial 
redshifts is supported 
by the decrease of the temperature
angular power spectrum at high multipoles 
and by an excess in the
TE cross-power spectrum on large angular scale
(multipoles $\ell \lsim 7$) with respect to models
with no-reionization or reionization with quite low ($\sim 0.05$) values
of $\tau$.
According to WMAP 1-yr data, 
the favourite values of $\tau$ are $\gsim 0.1$, 
with a current best fit of $\simeq 0.17$ based on the combination of WMAP with 
additional CMB anisotropy data at higher resolution and 
other kinds of astronomical observations (see, e.g.,
\cite{PBM_NA}).
On the other hand, 
the details and the physical 
explanation of the cosmological reionization 
are still unclear. 

Since the WMAP detection of the reionization
at substantial redshifts, many models have been 
proposed and/or reconsidered to account for the measure 
of $\tau$ (see, e.g., \cite{bean}).
Certainly, a significant contribution 
($\simeq 0.05-0.07$) to the value of $\tau$ 
derives from the ionization, 
likely associated to photon production
and hot gas ejection by primeval galaxies and quasars,
at relatively low redshifts ($z \lsim \zreio^{(2)} \simeq 5-7$) 
where direct observations of the Gunn-Peterson effect
in quasars probe a high ionization level of the intergalactic medium.
The remaining contribution ($\simeq 0.05-0.12$) to the observed value 
of $\tau$ should be provided by processes at higher redshifts.
Although uncertain because of the possible non-uniformity 
of the ionization of the intergalactic medium,
the indication of an increase of the Ly-$\alpha$ opacity 
for Gunn-Peterson tests toward some quasars at $z \sim 6$
\cite{becker, fan, fan2004}
combined to the large WMAP value of $\tau$
suggests the possibility of a twice reionization, the first one
at relevant redshifts, $\zreio^{(1)}\simeq 15$, possibly associated
to Population III stars \cite{cen2003}, or even at higher redshifts,
$\zreio \gsim some \times 10$, in models involving 
a relevant photon production by particle decay, 
matter-antimatter annihilation, or black-hole evaporation
and so on, followed by the subsequent 
reionization at $z \simeq \zreio^{(2)} \simeq 6$. 

The reionization imprints on the CMB 
can be divided in three categories: $i)$ decreasing of the power
of CMB temperature anisotropy at large multipoles
because of photon diffusion, $ii)$ increasing
of the power of CMB polarization and temperature-polarization
cross correlation anisotropy at all multipoles with relevant
features possibly more remarkable at low and middle multipoles 
according to the reionization epoch
because of the delay of the effective last scattering surface, 
$iii)$ generation of CMB Comptonization and free-free spectral distortions 
associated to the heating of the electron temperature of the 
intergalactic medium during the reionization epoch.
The imprints on CMB anisotropies are mainly dependent 
on the ionization history while 
CMB spectral distortions strongly depend 
also on the thermal history. 

    \section{Observational perspectives from next and future experiments}
    \label{sec:obs_anis}

The CMB anisotropy pattern is a single
realization of a stochastic process and therefore 
it may be different from the average over the ensemble 
of all possible realizations of the given (true) cosmological model 
with given parameters. 
This translates into the fact that the $a_{\ell m}$ coefficients
are random variables (possibly following a Gaussian distribution), at a 
given $\ell$, and therefore
their variance, $C_\ell$, is $\chi^2$ distributed with $2\ell +1$ degrees of
freedom. The relative variance $\delta C_\ell$ on $C_\ell$ is equal to 
$\sqrt{2/(2\ell+1)}$ and is quite relevant at low $\ell$s.
This is the so-called ``cosmic variance'' which limits the accuracy
of the comparison of observations with theoretical predictions.
Another similar variance in CMB anisotropy experiments 
is related to the sky coverage since the detailed CMB anisotropy 
statistical properties may depend on the considered sky patch.
This variance depends on the observed sky fraction, $f_{\rm sky}$. 
At multipoles larger than $few \times 10^2$ the most 
relevant uncertainties are related to the experiment resolution and 
sensitivity. All these terms contribute to the final uncertainty on the
$C_\ell$ according to \cite{knox}:
\begin{equation}
\frac{\delta C_\ell}{C_\ell} = \sqrt{\frac{2}{f_{\rm sky}(2\ell+1)}}\left[
1+\frac{A\sigma^2}{NC_\ell W_\ell}\right]\, ,
\label{fullvariance}
\end{equation} 
where $A$ is the size of the surveyed area, $\sigma$ is the rms noise per pixel,
$N$ is the total number of observed pixel, and $W_\ell$ is the beam window
function. For a symmetric Gaussian beam 
$W_\ell = {\rm exp}(-\ell(\ell+1)\sigma_{\rm B}^2)$ where
$\sigma_{\rm B} = {\rm FWHM}/\sqrt{8{\rm ln}2}$ 
defines the beam resolution.

The two instruments at cryogenic temperatures
on-board the ESA {\sc Planck}
satellite
\cite{tauber00} 
(see also J. Tauber 2004, {\it this Meeting}), 
the Low Frequency Instrument (LFI; \cite{lfi98};
see also N. Mandolesi 2004, {\it this Meeting})
based on differential radiometers  
and the High Frequency Instrument (HFI; \cite{hfi98})
based on state-of-art bolometers
at the focus of a 1.5~m aperture off-axis Gregorian telescope,
will measure the CMB angular power spectrum
with very high sensitivity up to multipoles $\ell \sim 1000-2000$
and an accurate control of the systematic effects.
The two instruments will cover a frequency range from 
30 to 857~GHz, necessary to accurately subtract the foreground
contamination (see, e.g., G. De~Zotti et al. 2004, {\it this Meeting}),
with a (FWHM) resolution from $\sim 33'$ to $5'$   
and a sensitivity from $\sim 15 \mu$K to the $\sim \mu$K level
in terms of antenna temperature on a square resolution element of with
side of $\sim 10'$. 

\begin{figure}
\vspace{-0.5cm}
\epsfig{figure=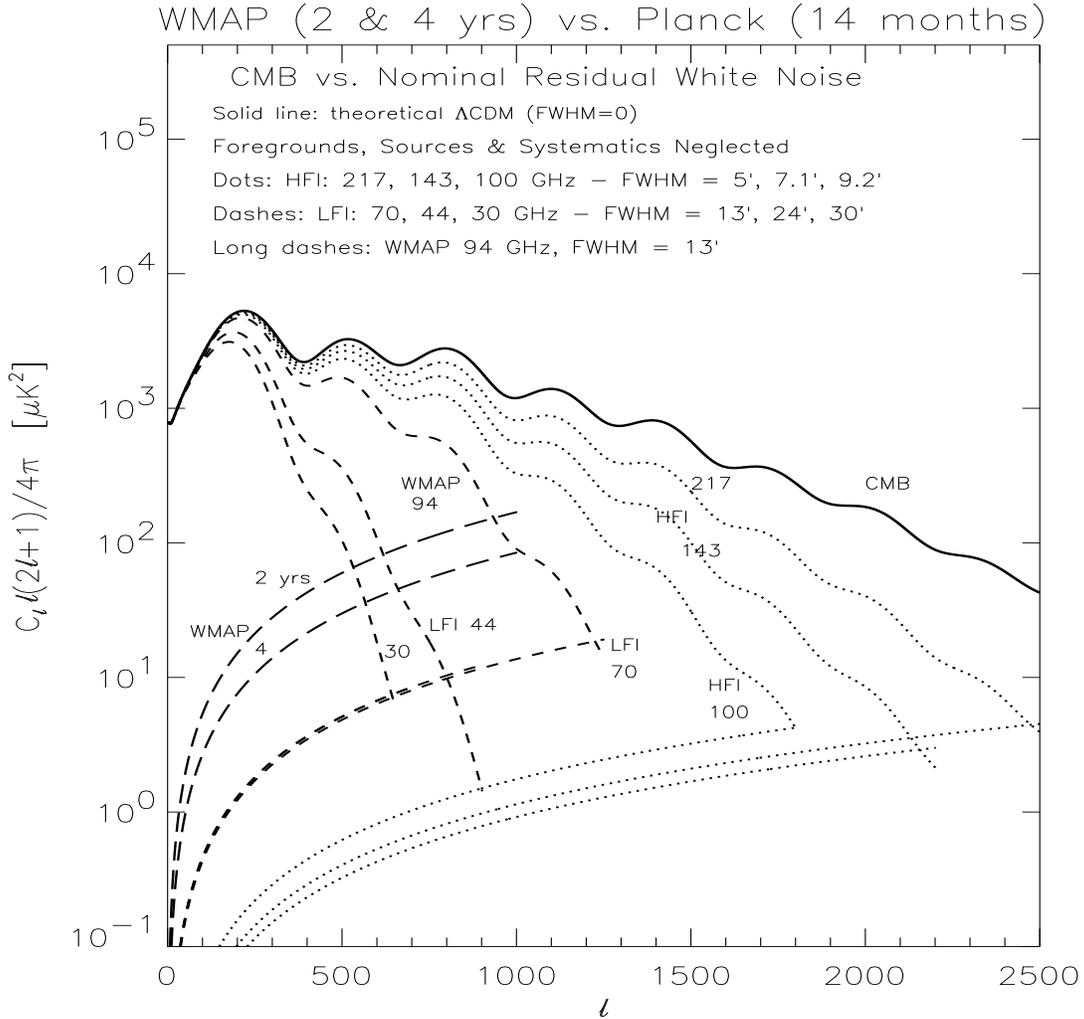,width=15cm,height=14cm}
\caption{Comparison between {\sc Planck} and WMAP resolution and
sensitivity. For each considered frequency
channel, the crossing between the CMB convolved angular power spectrum
and the residual instrumental white noise angular power spectrum
indicates the multipole value where the signal to noise ratio
($\ell$ by $\ell$) is close to unity.
Of course, binning the power spectrum on a suitable range of multipoles,
as possible because of the smooth 
variation of $C_\ell$ with $\ell$,
will allow to recover the CMB power spectrum with a good 
accuracy also at multipoles comparable that those 
corresponding to the crossings between the noise and CMB power 
spectra reported in the figure (for example 
for the {\sc Planck} channels at 70--100~GHz 
at $\ell \sim 1500$ a $\sim 3$\% binning over $\ell$ allows to 
reduce the uncertainty on the $C_\ell$ recovery to $\sim 15$\%).}
\label{planckvsmap}
\end{figure}

Fig.~\ref{planckvsmap} compares {\sc Planck} and 
WMAP
final performances
in terms of angular power spectrum recovery.
Note that, at similar frequencies, the better sensitivity of {\sc Planck}
with respect to WMAP is mainly due to the lower level of the {\sc Planck} 
instrumental noise achieved thanks to the better radiometer performance
assured by the lower system temperature while at higher frequencies
it derives from a combination of lower instrumental noise achieved thanks to 
high sensitivity bolometers and from the improving of resolution  
with the frequency for a given telescope size.
The CMB angular power spectrum is reported without beam smoothing and by taking into
account the beam window functions of several {\sc Planck} frequency
channels and of the highest WMAP frequency channel (which resolution is very close 
to that of the LFI 70~GHz channel). The corresponding angular power spectra
of the residual nominal white noise (i.e. after the subtraction of
the expectation value of the noise angular power spectrum) are also 
displayed. Similar considerations hold also for the measure of the polarization 
anisotropies (see, e.g., \cite{dsum02} for a summary of 
CMB polarization experiments). 
The sensitivity to the anisotropy measure of 
the Stokes parameters $Q$ and $U$ and of the polarization 
signal $P$ ($P=\sqrt{Q^2+U^2}$) is $\sim \sqrt{2} - 2$ times worse 
than that for the temperature anisotropy,
the exact value depending on to the detailed experimental strategy 
adopted to recover them from the combination of multi-beam 
data~\footnote{
For example, a ratio $\sqrt{2}$ can be easily obtained by taking into 
account that for differential radiometers the information from four receivers coupled to
two feeds can be combined to derive a measure of a single temperature 
anisotropy data or of two ($Q$ and $U$) polarization anisotropy data.}.
The low (of few--some~\%) polarization level predicted, and still only
approximately measured \cite{kovac2002}, for the CMB anisotropy clearly calls 
for high sensitivity measurements which are
less crucial for the temperature-polarization cross correlation
(TE-mode) already measured by WMAP.

\begin{figure}[t]
\vspace{-0.5cm}
\begin{center}
\epsfig{file=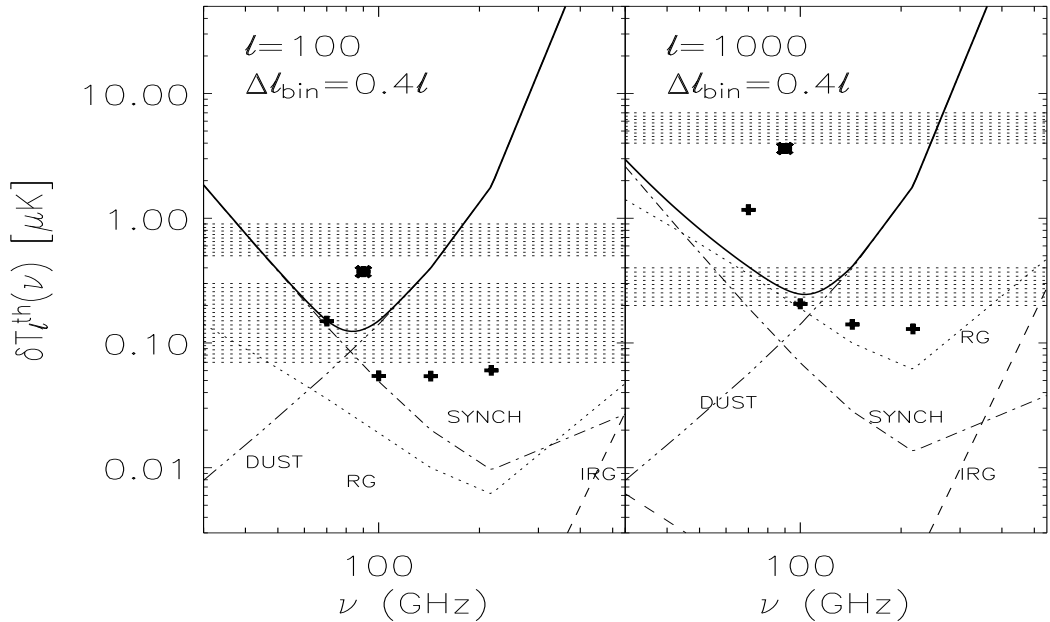,width=16.cm,height=14cm}
\end{center}
\vspace{-6.5cm}
\caption{Typical predicted ranges for the CMB 
polarization angular power spectra 
(E-mode, upper dotted regions; B-mode, lower dotted regions) 
including the lensing \cite{SZlensing}
compared to the 
foreground contamination (radiogalaxies (dots) contribution 
(see \cite{tucci}
for a recent detailed study);
the small contribution expected from infra-red galaxies (dashes); 
Galactic dust (three dots-dashes) and 
synchrotron (dot-dashes) emission \cite{bl02, dsum02})   
at two representative multipoles, as function of the 
frequency. 
We report the final sensitivity of WMAP at 94~GHz 
(asterisk) and of some {\sc Planck} frequency channels (crosses) binned
over a quite large range of multipoles.
}
\vspace{-0.2cm}
\label{cl_pol_nu}
\end{figure}

A quite accurate measure of the E-mode polarization 
is expected from the next WMAP data release, at least at degree scales,  
while the {\sc Planck} instruments will have a good sensitivity  
to the polarization E-mode angular power spectrum also at scales
of some arcminutes, the main limitation being represented 
by the foreground contamination, 
as shown in Fig.~\ref{cl_pol_nu}.

Thanks to the combination of temperature and polarization high quality data,
{\sc Planck} is expected to reduce the error bars in the recovery of the 
cosmological parameters at the level of
$few$~\% or better and the final accuracy 
on a wide set of cosmological parameters 
will be largely independent \cite{pbm01}
of the auxiliary information coming from
other classes of astronomical observations which will be
necessary to just remove some degeneracies intrinsic 
in the CMB information \cite{bondetal_deg}.
As an example, Fig.~3 compares the {\sc Planck}
sensitivity 
in constraining the Thomson optical depth
and the density contrast with 
the sensitivity of WMAP, possibly combined with the Ly-$\alpha$ forest 
information (from \cite{PBM_NA}).

The {\sc Planck} sensitivity and resolution will allow 
not only an extremely precise determination of the 
CMB angular power spectrum but also an accurate 
and multi-frequency imaging of the temperature anisotropy pattern 
on the whole sky and a quite accurate imaging of the polarization anisotropy
pattern at least on some sky areas of particular 
sensitivity~\footnote{
Analogously to the WMAP survey, the scanning strategies foreseen for 
{\sc Planck} will imply that the sky pixels on two areas (each of about 20--30
squared degrees) close to the ecliptic 
poles will be observed for a time significantly longer than the average 
resulting into a sensitivity significantly better 
(by about 5 times) than the average.}
by combining the multi-frequency information.
This is required for the study of the high order statistical
properties of the CMB anisotropy  
at high resolution, crucial to test some cosmological
scenarios predicting localized features (topological defects,
some inflationary models, ...).

\begin{figure}[t]
\vspace{-0.5cm}
\begin{center}
\epsfig{file=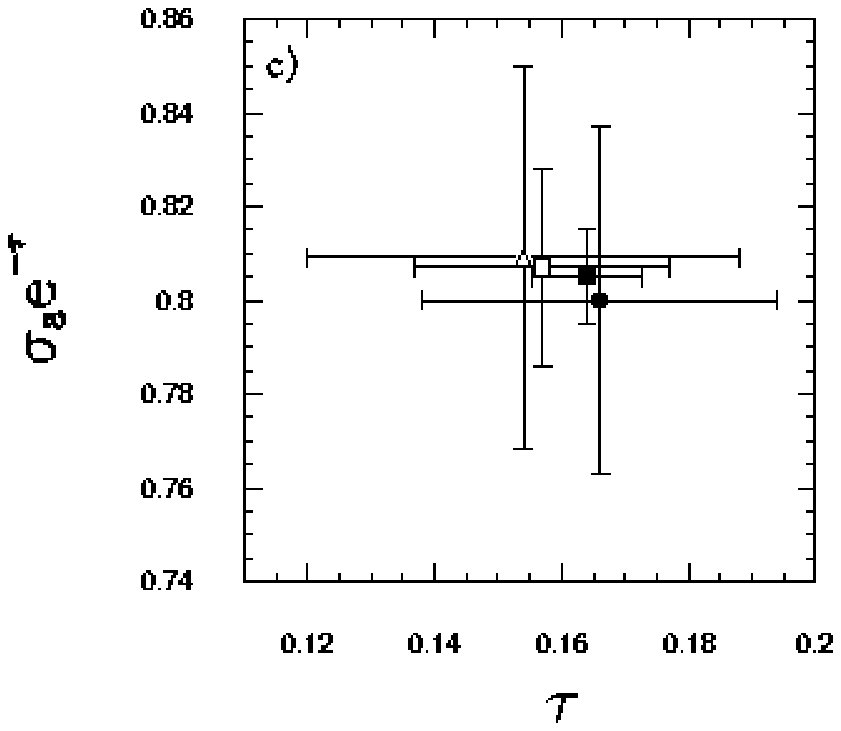,width=8.cm}
\end{center}
\caption{{\sc Planck}
sensitivity (filled square)
in constraining the Thomson optical depth
and the density contrast $\sigma_8$ compared with 
the sensitivity of WMAP 1-yr data (triangle), possibly combined with
the Ly-$\alpha$ forest information (filled circle),
and of WMAP 4-yr data (square). From \cite{PBM_NA}.}
\label{cosmo}
\end{figure}

The detection of the polarization B-mode angular power spectrum is within the 
{\sc Planck} capabilities at least over some suitable ranges of multipoles,
but its accurate and exhaustive measure requires a new generation
of dedicated experiments. 
As well known, 
the details of the reionization history strongly affect 
also the amplitude of the polarization B-mode angular power 
spectrum, which depends on the amplitude of tensorial
(gravitational waves) modes and, at small scales, on the lensing
effect.

The cosmic variance uncertainty
and the Galactic foreground contamination,
quite relevant at low multipoles, decreases at middle and 
high multipoles. The atmospheric emission is expected to be less crucial
for CMB polarization measurements than for anisotropy and spectrum ones.
Therefore, 
ground-based observations (\cite{hobson_tac}; see, e.g., the 
VSA project~\footnote{http://www.mrao.cam.ac.uk/telescopes/vsa/index.html})
appear of particular interest for obtaining precise information
at intermediate and high multipoles,
thanks to the high sensitivity that could be 
reached with
very long integration times and/or the use of a 
very large number of receivers.
On the other hand, the space is the favourite site for the accurate 
measure of the CMB polarization at low and intermediate multipoles
(see, e.g., the NASA Inflation 
Probe~\footnote{http://universe.nasa.gov/program/inflation.html}
of the Beyond Einstein program). 

Fig.~4 reports the uncertainty (including cosmic variance and 
instrument sensitivity and resolution)
in the recovery of the CMB polarization
power spectra as achievable with a $\sim$~yr full sky space 
mission discussed in the context of 
the call of mission themes of the ESA Cosmic Vision 2015-2025,
by using the state 
of the art radiometers (right panel) and bolometers (left panel)
at the focus of a set of four $\sim 60$~cm telescopes.

The comparison with Fig.~5 shows that, as long as foreground
polarization anisotropies (and, in particular, the Galactic
contamination) are accurately known and subtracted, 
the sensitivity levels presented above will allow
to accurately measure also the B-mode or, depending
on its intrinsic level, to set significant constraints on it.

\begin{figure}[t]
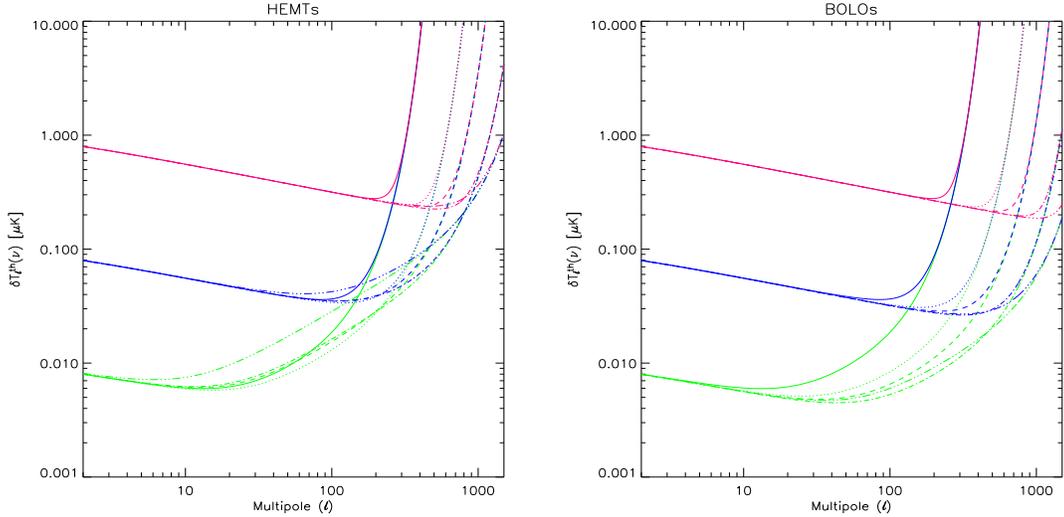

\begin{center}
\begin{tabular}{cc}
\includegraphics[width=7cm,height=7cm]{sensitivity_clpol_reno.ps_pages1}&
\includegraphics[width=7cm,height=7cm]{sensitivity_clpol_reno.ps_pages2}
\end{tabular}
\end{center}
\caption{Residual uncertainty in the knowledge of the polarization
angular power spectrum (in terms of 
$\delta T = (\ell(\ell+1)C_\ell/2\pi)^{0.5}$; thermodynamic
temperatures are considered)
including cosmic variance and detector
sensitivity and resolution as in principle achievable by the 
next generation of CMB space missions.
Right panel: the case of radiometer technology
(typical sensitivity of $\simeq 0.1-0.25 \mu$K on a pixel of $30'$
side).
Left panel: the case of bolometer technology
(typical sensitivity of $\simeq 0.04-0.12 \mu$K on a pixel of $30'$
side).
In each panel, the different lines refer to frequency
channels at 32, 64, 94, 143, 217 GHz; at increasing frequency 
the error at high multipoles decreases because
of the resolution increasing 
(from instance 1.09$^\circ$ to 0.16$^\circ$ (FWHM))
by using a given telescopes size.  
The cosmic variance has been computed by assuming, for simplicity, 
a constant polarization anisotropy level of 1, 0.1, and 0.01 $\mu$K
in terms of $\delta T$ (lines from the top to the
bottom at low multipoles). Full sky coverage is assumed.}
\label{pol_sensi}
\end{figure}

The possibility to significantly improve our knowledge 
of the CMB spectrum with respect to the current observational
status, in which the COBE/FIRAS data play the major role
in constraining the CMB spectral distortions,
has been recently addressed.
In particular, 
two space mission proposals, 
the DIMES experiment \cite{KOG96},
designed to reach an accuracy close to that of FIRAS
but at centimeter wavelengths, and 
the FIRAS~II experiment \cite{FM02}
which will allow a sensitivity improvement by a factor $\sim
100$ with respect to FIRAS,
open the new perspective to 
detect CMB spectral distortion and not only to set constraints
on them.

DIMES 
\cite{KOG96} is a space mission submitted to the NASA in
1995, designed to measure very accurately the CMB spectrum at wavelengths
in the range $\simeq 0.33 - 15$~cm \cite{KOG96}.
DIMES will compare the spectrum of each $\simeq 10$ degree pixel on the 
sky to a precisely known blackbody to precision of $\sim 0.1$~mK,
close to that of FIRAS.
The set of receivers is given
from cryogenic radiometers
with instrument emission cooled to 2.7~K
operating at six frequency bands
about 2, 4, 6, 10, 30 and 90~GHz using a single external blackbody
calibration target common to all channels to minimize the calibration
uncertainty.
The DIMES design is driven by the
need to reduce or eliminate systematic errors from instrumental artifacts.
The DIMES sensitivity represents an
improvement by a factor better than 300 with respect to
previous measurements at centimeter wavelengths allowing a significant
improvement of our knowledge on early dissipation processes and
free-free distortions \cite{BS03a}.

\begin{figure}[t]
\vspace{-0.5cm}
\epsfig{file=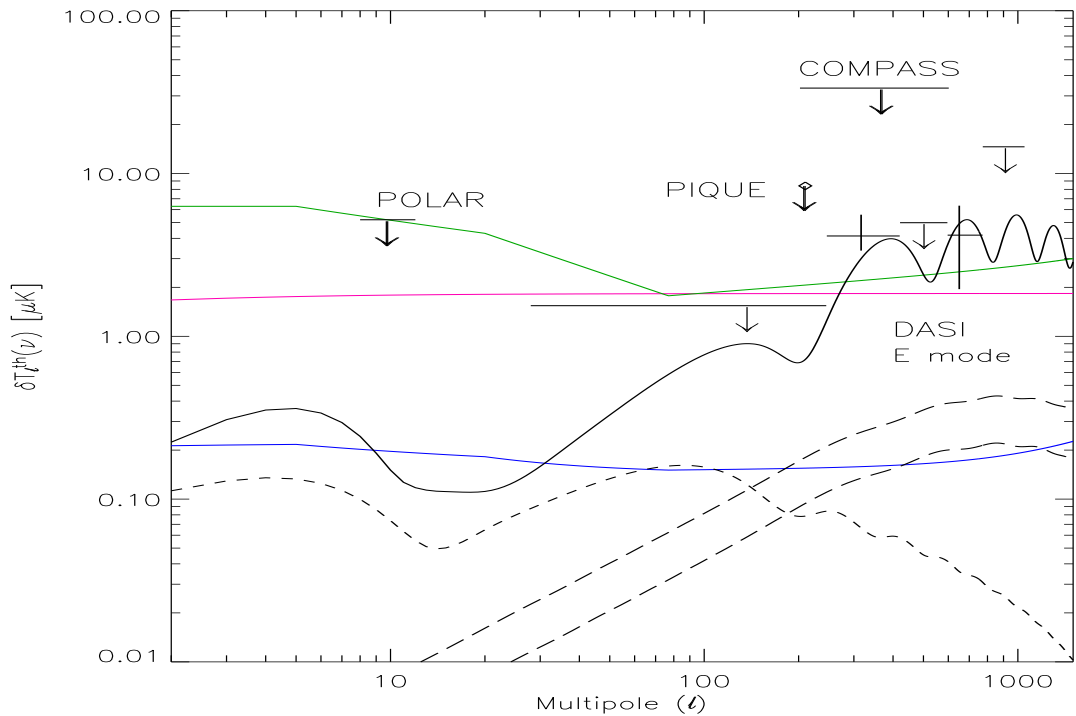,width=15cm,height=10cm}
\caption{CMB polarization angular power spectra (E-mode: black
solid line; B-mode: black dashes; B-mode from lensing:
black long dashes) compared with reasonable predictions
for the overall (Galactic plus extragalatic) foreground
polarized angular power spectrum at 30 (green line),
100 (blue line) and 217~GHz (red line). The few recent upper 
limits and measures are reported for comparison 
\cite{polar, pique, compass, kovac2002}.}
\label{clpolfore}
\end{figure}
\vspace{8cm}

Fixsen and Mather \cite{FM02} described the fundamental guidelines
to significantly improve CMB spectrum measures at $\lambda \lsim 1$~cm.
A great reduction of the residual noise of cosmic rays, dominating
the noise of the FIRAS instrument, can be obtained by eliminating
the data on-board co-add process or applying deglitching before co-adding,
by reducing the size of the detectors and by using
``spiderweb'' bolometers or antenna-coupled micro-bolometers.
They are expected to show a very low noise when cooled below 1~K
while RuO  sensors can reduce to 0.1~mK the read noise of thermometers.
The Lagrangian point L2 of the Earth-Sun system is, of course, the
favorite ``site'' for FIRAS~II. Also, the calibration can be
improved by order of magnitudes
with respect to that of FIRAS by reducing the contribution to
the calibrator  reflectance  of light from the diffraction
at the junction between the calibrator and the horn.
A complete symmetrical construction of the instrument is
recommended. This allows the cross-check between calibrators
and between calibrators and the sky and to realize
``an end-to-end calibration and performance test before launch''.
According to the authors, FIRAS~II can be designed to have
a frequency coverage from 60 to 3600~GHz
(i.e. from 5~mm to 83~$\mu$m) with a spectral
resolution $\nu / \Delta \nu < 200$ and sensitivity
in each channel about 100 times better than that of FIRAS.

The combination of these two spectrum experiments 
open, at least in principle, the perspectives to detect and 
possibly measure spectral distortions imprinted by energy 
dissipations about 100 times smaller than those corresponding
to the current upper limits set by FIRAS \cite{BS03b}.

\section{Reionization phenomenological models}
\label{subsec:reion_phen_mod_pol}

The reionization process 
can be described phenomenologically in terms of injection of
additional ionizing photons 
\cite{psh, dn, dnnn}.
The ionization
fraction of matter, $x_e=n_e/\overline{n}$, can be obtained
from the balance between the processes of recombination and ionization:
\begin{equation}
\frac{dx_e}{dt}=-\alpha_{\rm rec}(T) \nb x^2_e +
\varepsilon_{i}(z) (1-x_e) {{H}}(z),
\label{eq3}
\end{equation}
where 
$\nb(z)$ is and the mean baryonic density at the redshift $z$,
$T$ is the temperature of the plasma, 
$\alpha_{\rm rec}(T)\simeq 4\times 10^{-13} \left(T/10^4
K \right)^{-0.6} {\rm s}^{-1}{\rm cm}^{-3}$ is the recombination coefficient, 
and $\varepsilon_{i}(z)$ is the efficiency of the  
ionizing photon production giving 
the ionizing photon production rate 
${dn_i}/{dt}=\varepsilon_{i}(z) \nb(z) {{H}}(z) \, ;$
$H(z)=1/t_{exp}$, where $t_{exp}=a/(da/dt)$
is the cosmic expansion time, $a$ being here the cosmic scale
factor.

\noindent
The choice of the function $\varepsilon_{i}(z)$
allows to model the ionization history
also in the presence of extra sources of ionizing photons
(e.g. Ly-$\alpha$ and ionization photons from 
primavel stars, galaxies, and active galaxies, or from 
primordial black hole decays, electromagnetic cascades from particle 
release from topological defects or decay of super heavy dark matter, 
...).  

Assuming equilibrium 
between the recombination and the ionization process, 
for any given history of the plasma temperature 
the evolution of the ionization fraction
is given by the solution of the simple second order equation
obtained by the equality ${dx_e}/{dt}=0$. 

\subsection{Late processes}

For late processes, a simple expression  
describing a first reionization at relevant redshifts
followed by a second reionization, 
possibly mimicking the model by Cen \cite{cen2003}, 
has been proposed~\cite{nc03} in the form:

\begin{equation}
\varepsilon_{i}(z)= \varepsilon_{0} \exp
\left[-\frac{(z-\zreio^{(1)})^2}{(\Delta z_1)^2}\right]
+\varepsilon_{1}(1+z)^{-m}\Theta(\zreio^{(1)}-z) \, ;
\label{eq5.0}
\end{equation}
here $\varepsilon_{0}$,
$\zreio^{(1)}$, and 
$\Delta z_1\ll \zreio^{(1)}$ are free
parameters describing the history
of the first epoch of reionization
which significantly decreases at $z > \zreio^{(1)}$;
$\varepsilon_{1}$, $m$, and (again)
$\zreio^{(1)}$ are free
parameters describing the history
of the second epoch of reionization
resulting into a increasing of 
$\varepsilon_{i}(z)$ with the time,
being $\Theta(x)$  the step function.

Although quite weakly in this modelization, because of the dependence
on the temperature as a power of $\simeq -0.6$
of the recombination coefficient, 
the thermal history influences the ionization fraction
evolution. By using the
matter temperature evolution
$T(z)\simeq 270 \left(1+z/100 \right)^2 {\rm K}$
and modifying the ionization history 
in the {\sc cmbfast} code, 
the CMB angular power spectrum, in both temperature
and polarization, has been computed \cite{nc03} for some representative choices 
of the above free parameters.
As shown by the authors
for the case of the E-mode polarization power spectrum,
the differences between different late reionization scenarios
associated to very different parameters are of particular interest
at relatively low multipoles. 
Provided that the contribution from the Galactic foregrounds,
particularly relevant at these multipoles, could be accurately
modeled, the difference between a scenario involving a single 
reionization at $z \simeq \zreio^{(2)}$ and 
scenarios involving a twice reionization could be
detected by the {\sc Planck} polarization accuracy.


\begin{figure}[t]
\vspace{-1.5cm}
\begin{center}
\epsfig{file=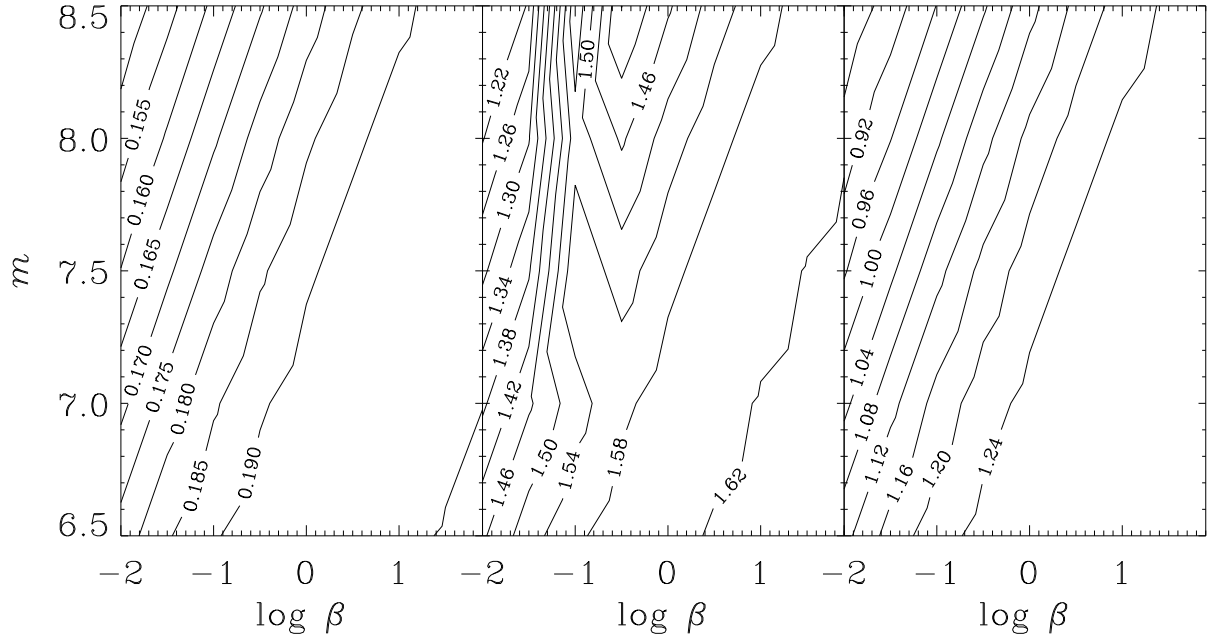,width=16.cm}
\end{center}
\vspace{-7cm}
\caption{Contour plots of the optical depth and 
of the fractional injected energy associated to the Comptonization parameter
as functions of $\beta$ and $m$ 
in the case of late reionization processes.
Left panel: $\tau$
evaluated by integrating up to $z=0$.
Middle and right panels: $\Delta \epsilon / \epsilon_i \simeq 4u$ 
(in units of $10^{-6}$)
computed by integrating up to $z=0$ and up to $z=5$,
respectively.}
\label{late_all}
\end{figure}

The relevance of the detection of spectral distortions 
has been partially renewed 
\cite{KOG03}
by the WMAP satellite discovery of a
reionization phase at relevant redshifts.

The computation of the CMB spectral distortions associated to 
the reionization parametric model presented here
is strictly related to the detailed description 
of the thermal history.

For these kinds of late processes,   
we assume here a matter thermal history similar to that reported 
by Cen \cite{cen2003}: a matter temperature $T$ in approximate thermal equilibrium with the 
radiation field (i.e. equal to $T_0 (1+z)$) at $z \gsim 27$;
a linear dependence of ${\rm log} T$ on $z$ at subsequent times
to reach a temperature $T^{(1)}$ (assumed here $1.5 \times 10^4$~K
for numerical estimates) at $z=\zreio^{(1)}$, kept then constant 
up to $z=\zreio^{(2)}$ when it rapidly increases up to
a temperature $T^{(2)}$ (assumed here $2 \times 10^4$~K
for numerical estimates) kept then constant up to low redshifts.   

We have implemented this thermal history in the 
equilibrium evolution of the 
ionization fraction and then used 
these thermal and ionization histories to 
compute the Comptonization and free-free distortions
as described in \cite{BDD95} 
by generalizing the code to include also the case of a $\Lambda$CDM model.

In Fig.~\ref{late_all} we report our 
contour plot results for
the optical depth $\tau$ and the fractional energy injection
associated to the Comptonization distortion 
as functions of the free parameters   
$\beta=\varepsilon_{1}/10^9$ and $m$,
by separately showing for $\Delta \epsilon / \epsilon_i$ 
the results of the integration up to
$z=0$ and up to $z=5$ (of course, the difference 
between these results gives the fractional energy injected
in the plasma at $z \le 5$). For simplicity, we report here
the results for an interval of $\beta$ and $m$ producing values of $\tau$ 
close to the current WMAP best fit
\cite{kogutetal03}. 
Although a detailed computation requires to take into account
the dependence of the ionization history on the matter temperature 
evolution, these results can be approximately 
rescaled to different values of $T^{(1)}$ and $T^{(2)}$ by considering 
the linear dependence of the Comptonization parameter on $T$. 
These predicted Comptonization
distortions could detected and possibly measured by 
an experiment with a sensitivity like that of FIRAS~II.
Analogously to the case of the polarization anisotropy signatures,
the difference between the Comptonization distortions by 
models involving a twice reionization (or a 
continuous reionization starting at 
$z \simeq \zreio^{(1)}$)
and models with a single reionization at $z \simeq \zreio^{(2)}$
is comparable or above the FIRAS~II sensitivity.
The free-free distortion parameter $y_B$ results to be too small,
well below the DIMES sensitivity, 
and is not reported here~\footnote{On the contrary, note that
several specific physical models predict free-free distortion
levels larger than those found in this simple modelization
and clearly observable by a DIMES-like experiments
(see, e.g., \cite{BS03b} and references therein).}.

\subsection{High redshift processes}

Models involving a substantial reionization at redshifts, 
$z \simeq \zreio$, much higher than $\zreio^{(1)} \sim 15$
can not be excluded by current data.
A Gaussian parametric form 

\begin{equation}
\varepsilon_{i}(z)= \xi
\exp \left[-\frac{(z-\zreio)^2}{(\Delta z)^2}\right] \, ,
\label{eqc9}
\end{equation}

\noindent
quite similar to that assumed to describe 
the first reionization epoch for late processes,
can be exploited in this context \cite{nc03};
again $\xi$, $\zreio$, and 
$\Delta z$ are free
parameters describing the history
of this high redshift reionization scenario.
Assuming $\Delta z \ll \zreio$ implies the choice 
of a peak-like model.
Again, it is necessary to add the evolution of the matter 
temperature which should peak at typical values 
of $T_p \approx (1-2) \times 10^4$~K. 
In \cite{nc03} it is assumed an extremely short heating phase 
($\delta$-like) about $\zreio$. After the end of the heating phase
the matter temperature is determined by the usual equation
for the electron temperature evolution, dominated, at high redshifts,
by the Compton cooling term. Therefore a set of 
two differential equations, one  
for the ionization fraction and the other for the matter temperature,
describe the problem~\footnote{Simple approximate analytical solution 
can be used in the limit $\xi \ll 1$. We find also that 
the the system can be easily solved through a numerical integration. 
A simple backward differential scheme with a very small time 
step compared to the other relevant timescales does not introduce 
relevant errors in this context.}. 
This modified ionization history has been 
included in the {\sc cmbfast} code to  
computed the CMB angular power spectrum, in both temperature
and polarization, for some representative choices 
of the above free parameters.
As shown by the authors
for the E-mode polarization angular power spectrum,
the differences between different early reionization scenarios
associated to different parameters 
and between them and a scenario involving a single late
reionization at $z \simeq \zreio^{(2)}$
are relevant at low and also at middle multipoles, 
$\ell \sim few \times (10-100)$. 
This is of particular interest in the light of the 
{\sc Planck} polarization accuracy in the 
multipole region of the CMB acoustic peaks
where the contribution from the Galactic foregrounds
is expected to significantly decrease with respect to the level 
it has at low multipoles.


%
\begin{figure}
\vspace{-1.5cm}
\begin{center}
\epsfig{file=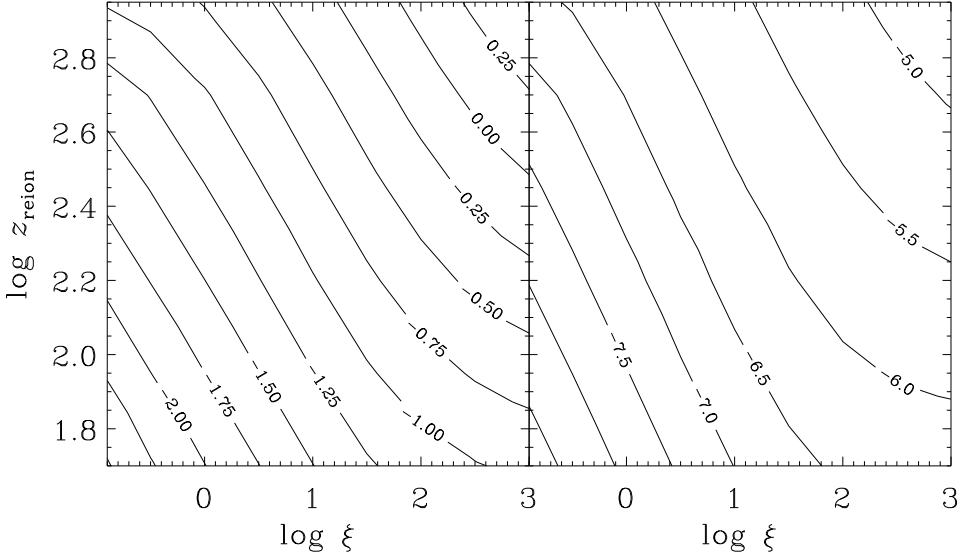,width=14.cm}
\end{center}
\vspace{-7cm}
\caption{Contour plots of the optical depth, $\tau$,
(left panel) 
and of the fractional energy injection 
associated to the Comptonization parameter, 
$\Delta \epsilon / \epsilon_i \simeq 4u$, 
(right panel)
as functions of $\xi$ and $\zreio$
in the case of early peak-like reionization processes (logarithmic scale).}
\label{early_all}
\end{figure}

For the study of the spectral distortions 
associated to early peak-like reionization processes,
we have numerically integrated the differential equations 
for the ionization fraction and the matter temperature
as described above.
Differently from 
the $\delta$-like assumption 
for the matter temperature during the heating phase
about $\zreio$ (that is not particularly critical 
for polarization anisotropy considerations)
we adopt here a Gaussian parametric form 

\begin{equation}
T(z)= T_p \exp \left[-\frac{(z-\zreio)^2}{(\Delta z)^2}\right] \,  
\label{eqc9_T}
\end{equation}

\noindent
for the matter temperature during the heating phase
(within a redshift interval $\simeq \pm 3 \Delta z$
about $\zreio$) and then assume the usual 
temperature evolution equation, dominated by the  
Compton cooling term, at later epochs
($T_p = 1.5 \times 10^4$~K 
and $\Delta z = 0.025 \zreio$
are adopted here for numerical estimates).
It is in fact quite reasonable to assume a similar
time dependence for both the efficiency of the  
ionizing photon production, $\varepsilon_{i}(z)$, 
and the matter temperature during the active phase.
We assume as initial conditions 
at the beginning of the heating/ionization phase
the thermal equilibrium between matter and radiation
and the residual ionization fraction obtained 
at the considered redshift from the standard 
recombination.

Finally, we have implemented the thermal and ionization history
as described above in the code \cite{BDD95} 
for the computation of the Comptonization and free-free distortions.

In Fig.~\ref{early_all} we report our 
results in terms of contour plots
for the optical depth $\tau$ and the fractional energy injection
associated to the Comptonization distortion 
as functions of the free parameters   
$\xi$ and $\zreio$. These results can be 
rescaled to different choices of 
$T_p$ and $\Delta z$ by using the approximate
proportionality relations between  
$\tau$ and $\Delta z$  and between
$u$ and $T_p \Delta z$~\footnote{In reality, 
for the lowest considered values of $\xi$ ($\xi \approx 1$) 
we find an increasing 
of the Comptonization distortion with 
$\Delta z$ larger by a factor $\approx 1.5-3$
than that suggested by the above proportionality and
related to the different initial conditions 
(the process starts at earlier times) 
and to the different coupling between 
the ionization fraction and the matter temperature.
We find that this holds also for the free-free
distortion parameter $y_B$.}.
Fig.~\ref{early_all} shows that a large region 
of the $\xi$ and $\zreio$ parameter space
can be clearly ruled out by current WMAP data
because of the violation of the limits 
on the optical depth. It is also remarkable
that in the permitted region of $\xi$ and $\zreio$ 
(corresponding to $\tau \sim 0.1$) the model 
predicts values of fractional injected energy 
$\Delta \epsilon / \epsilon_i \simeq 4u \sim 10^{-6}$,
again in principle measurable with the FIRAS~II sensitivity.
We find again values of free-free distortion parameter $y_B$ too small
with respect to the DIMES sensitivity~\footnote{By considering
higher matter temperature values (up to $6 \times 10^4$~K),
we find only a weak dependence of $y_B$ on $T_p$ 
because the typical decreasing of $y_B$ with $T_p$ is approximately
compensated by the increasing of the ionization fraction
because of the decreasing of the recombination efficiency.}. 
 
To summarize, since it is plausible 
to assume peak matter temperatures $\gsim (1-2)\times 10^4$~K 
to have an efficient (late or early) reionization,
values of 
$\Delta \epsilon / \epsilon_i \simeq 4u \sim (1-2) \times 10^{-6}$
should be considered as typical (lower limit) predictions 
for the Comptonization distortion associated to reionization 
scenarios compatible with the WMAP results
\cite{buriganaetal04rev}.

\section{Inhomogeneous reionization in the presence of massive neutrinos}

\begin{figure}
\vspace{-0.5cm}
\epsfig{file=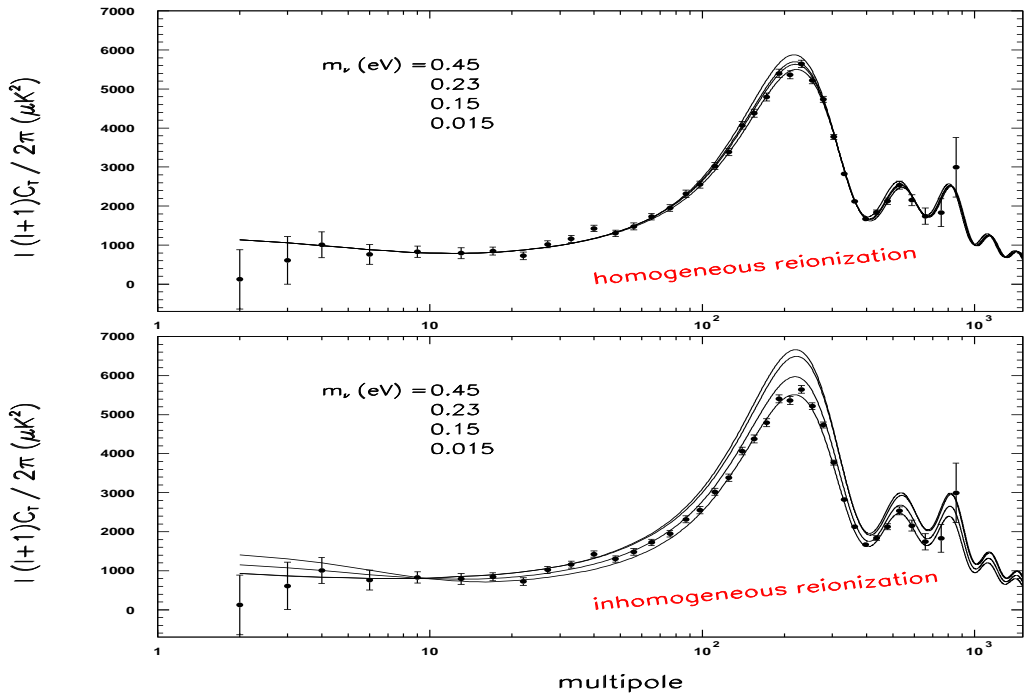,width=15.cm,height=10.cm}
\caption{The CMB anisotropy power spectra for different neutrino masses
considering (bottom panel) or not (top panel) the 
inhomogeneities of the reionization process.}
\label{clstar}
\end{figure}

As a specific example, we have investigated the role of a HDM component in 
the form of three massive neutrino flavors in the context of 
reionization scenarios.
Assuming a flat background cosmology
described by the best fit power law $\Lambda$CDM model with WMAP data 
($\Omega_bh^2=0.024$, $\Omega_mh^2=0.14$, $h=0.72$), 
we analyze the role of the neutrino mass for the properties of the gas 
in the intergalactic medium (IGM). We find that the temporal 
evolution of the hydrogen and helium ionization fractions is 
sensitive to the neutrino mass, with relevant implications for the CMB
anisotropy and polarization angular power spectra.

\begin{figure}
\vspace{-0.5cm}
\epsfig{file=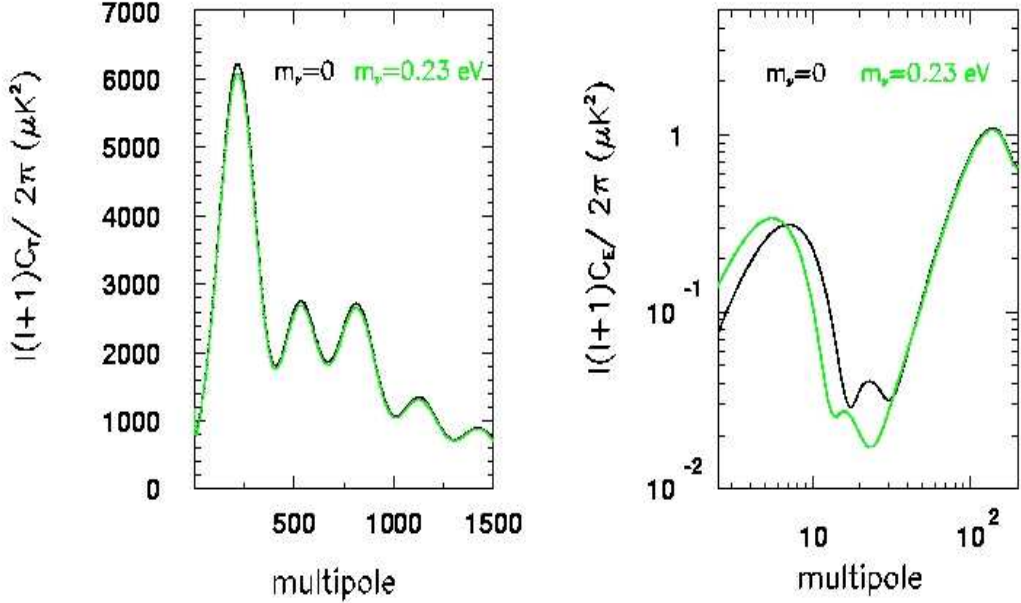,width=15.cm,height=10.cm}
\caption{CMB temperature (left panel) and polarization (E-mode, right 
panel) anisotropy power
spectra obtained for the above reionization model (including
inhomogeneities) by considering or not the neutrino contribution
for a specific value (25) of the reionization redshift.
Note the sensitivity of polarization to the neutrino mass.}
\vspace{1cm}
\label{cl}
\end{figure}

The reionization is assumed to be caused by the ionizing photons produced
in star-forming galaxies and quasars. 
This process depends on the evolution of the background density 
field that determines the formation rate of the bounded objects, 
the gas properties in the IGM and their feedback relation.
At the redshift of collapse, the fraction of the mass of the gas in
the virilized halos can be obtained if the probability distribution 
function of the gas overdensity is known.
The temperature-density relation and the virial mass-temperature relation  
determines the connection between the gas density and the matter density 
at the corresponding scales \cite{PBM_NA}.
During and after recombination,
neutrinos with masses in eV range can have significant interactions 
with photons, baryons and cold dark matter particles via gravity only. 
The neutrino phase-space density
is constrained by the Tremaine-Gunn criterion \cite{tg79} 
that puts limits on the neutrino energy-density inside bounded objects. 
Neutrinos cannot cluster via gravitational instability on scales 
below the free-streaming scale.
The free-streaming distance determines the neutrino energy-density 
distribution inside the gravitationally collapsed objects \cite{pbm02}, 
the temperature-density relation of the gas in the IGM and the hydrogen 
and helium ionization fractions.

%
\begin{figure}
\vspace{-0.5cm}
\epsfig{file=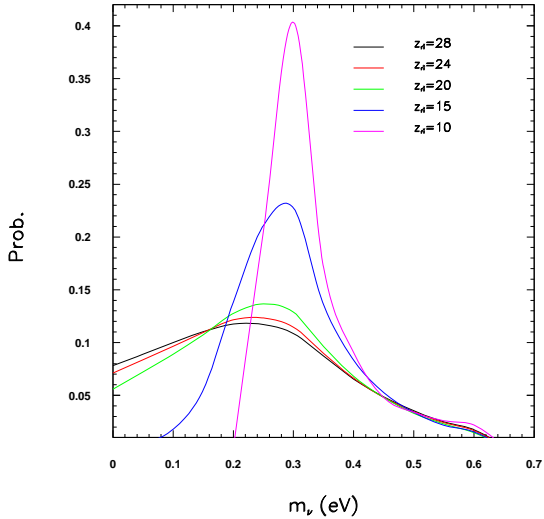,width=8.cm}
\caption{The probability density function of the neutrino
masses (69\% CL) as could be estimated by the {\sc Planck}
mission, using the CMB polarization measurements,
for different reionization redshifts
(fiducial model with $m_\nu = 0.23$~eV).}
\vspace{1cm}
\label{znu}
\end{figure}

In Fig.~8 we show the WMAP data compared with the CMB anisotropy 
power spectra obtained for different neutrino masses
with (bottom) and without (top) considering the 
inhomogeneities of the reionization process.
Fig.~9 compares the CMB temperature and polarization (E-mode) 
anisotropy power spectra obtained for the above reionization model 
(including inhomogeneities) by considering or not the neutrino contribution
for a specific value of the reionization redshift.
Clearly, polarization information greatly helps to distinguish
between the two cases.

Finally, in Fig.~10 we show how in this specific reionization model
the neutrino mass could in principle be constrained by {\sc Planck} 
depending on the assumed reionization redshift.
Note the improvement of the sensitivity to neutrino mass
with the decreasing of the reionization redshift,
as intuitive is because of the corresponding increase of the 
neutrino mass role.

\section{Conclusion}

The WMAP detection of a high redshift reionization 
calls for a better understanding of the physics of the 
reionization process. Different models predicts different imprints
on the CMB in both temperature and polarization anisotropies
while the presence of late spectral distortions 
(Comptonization like, with 
$\Delta \epsilon / \epsilon_i \sim {\rm some} \times 10^{-6}$, and also 
free-free distortions
with amplitudes strongly dependent on the specific considered 
models) seems unavoidable for reasonable thermal histories
in the context of reionization scenarios compatible with 
WMAP results.

Accurate measures of the CMB properties, such as those expected 
by the forthcoming {\sc Planck} satellite and by  
future experiments will offer the opportunity to 
constrain the ionization and thermal history 
of the universe at moderate and high redshifts.

\vspace*{1 cm}

\noindent{\bf Acknowledgements}

\vspace*{0.3 cm}

\noindent
We thank L.~Danese,
L.~La~Porta, and L.~Toffolatti for collaborations and discussions.
We also thank L.-Y. Chiang, B.~Ciardi, A.~Ferrara, and P.~Naselsky
for useful conversations on the reionization history.
The use of the {\sc cmbfast} code is acknowledged.
Some of the results in this paper have been derived applying the 
HEALPix~\footnote{http://www.eso.org/science/healpix/}
({\it Hierarchical Equal Area and IsoLatitude Pixelization
of the Sphere}) by G\`orski et al.~\cite{ref_healpix}) 
to the WMAP 1-yr data products.
Some of the calculations presented here have been
carried out on an alpha digital unix machine 
at the IFP/CNR in Milano by using some NAG integration codes.
The staff of the LFI DPC in Trieste, where
some simulations have been carried out, is acknowledged.


\end{document}